\documentclass[twocolumn,showpacs,preprintnumbers,amsfonts,amsmath,amssymb,prl,floatfix]{revtex4}

\usepackage{graphicx}
\usepackage{amssymb}
\usepackage{amsmath}

\newcommand{\ket}[1]{|#1\rangle}
\newcommand{\bra}[1]{\langle #1|}


\begin{document}

\title{Prospects for a mHz-linewidth laser}

\author{D. Meiser}
\author{Jun Ye}
\author{D. R. Carlson}
\author{M. J. Holland}
\affiliation{JILA, National Institute of Standards and Technology and University of Colorado\\
  Department of Physics, University of Colorado, Boulder, CO 80309-0440, USA}

\date{\today}

\begin{abstract}
We propose a new light source based on having alkaline-earth atoms in an
optical lattice collectively emit photons on an ultra-narrow clock transition
into the mode of a high Q-resonator. The resultant optical radiation has an
extremely narrow linewidth in the mHz range, even smaller than that of the
clock transition itself due to collective effects. A power level of order
$10^{-12}W$ is possible, sufficient for phase-locking a slave optical local
oscillator. Realizing this light source has the potential to improve the
stability of the best clocks by two orders of magnitude.
\end{abstract}

\maketitle

Time and frequencies are the quantities that we can measure with the highest
accuracy by far. From this fact derives the importance of clocks and frequency
standards for many applications in technology and fundamental
science. Some applications directly relying on atomic clocks are GPS,
synchronization of data and communication networks, precise measurements of the
gravitational potential of the earth, radio astronomy, tests of theories of
gravity, and tests of the fundamental laws of physics.

With the advent of octave spanning optical frequency combs
\cite{Hansch:comb,CundiffYe:comb} it has become feasible to use atomic
transitions in the optical domain to build atomic clocks. Optical clocks based
on ions \cite{T.Rosenband03282008EtAl} and ultracold neutral atoms
confined in optical lattices
\cite{ADLudlow03282008EtAl} have
recently demonstrated a precision of about 1 part in $10^{15}$ at one second
and a total fractional uncertainty of $10^{-16}$~\cite{ADLudlow03282008EtAl} or
below~\cite{T.Rosenband03282008EtAl}, surpassing the primary cesium microwave standards
\cite{NISTCsEtAl,SyrteCsEtAl}.

The state-of-the-art optical atomic clocks do not achieve the full stability
that is in principle afforded by the atomic transitions on which they are
founded. These transitions could have natural line-$Q$s of order $10^{18}$,
exceeding the fractional stability of the clocks by a factor of
$\sim 100$. The main obstacle that prevents us from reaping the full benefit of
the ultra-narrow clock transitions is the linewidth of the lasers used to
interrogate these transitions. These lasers are stabilized against carefully
designed passive high-$Q$ cavities and achieve linewidths $<1$ Hz,
making them the most stable coherent sources of radiation. It is mainly the
thermal noise of the reference cavity mirrors that prevent a further
linewidth reduction \cite{LudlowEtAl:Laser} and substantially reducing this
noise is hard \cite{Kimble:Thermalnoise}.

An elegant solution to these problems would be to directly extract light
emitted from the ultra-narrow clock transition
\cite{Chen:ChineseScienceBulletin}. That light could then be used as an optical
phase reference, circumventing the need for an ultra stable reference cavity.
Unfortunately, the fluorescence light emitted on a clock transition is too weak
for practical applications. For instance, for $10^6$ fully inverted ${}^{87}$Sr
atoms the power of the spontaneously emitted light is of the order of
$10^{-16}$~W.

\begin{figure}[h]
  \includegraphics[height=2.9cm,width=3.2cm]{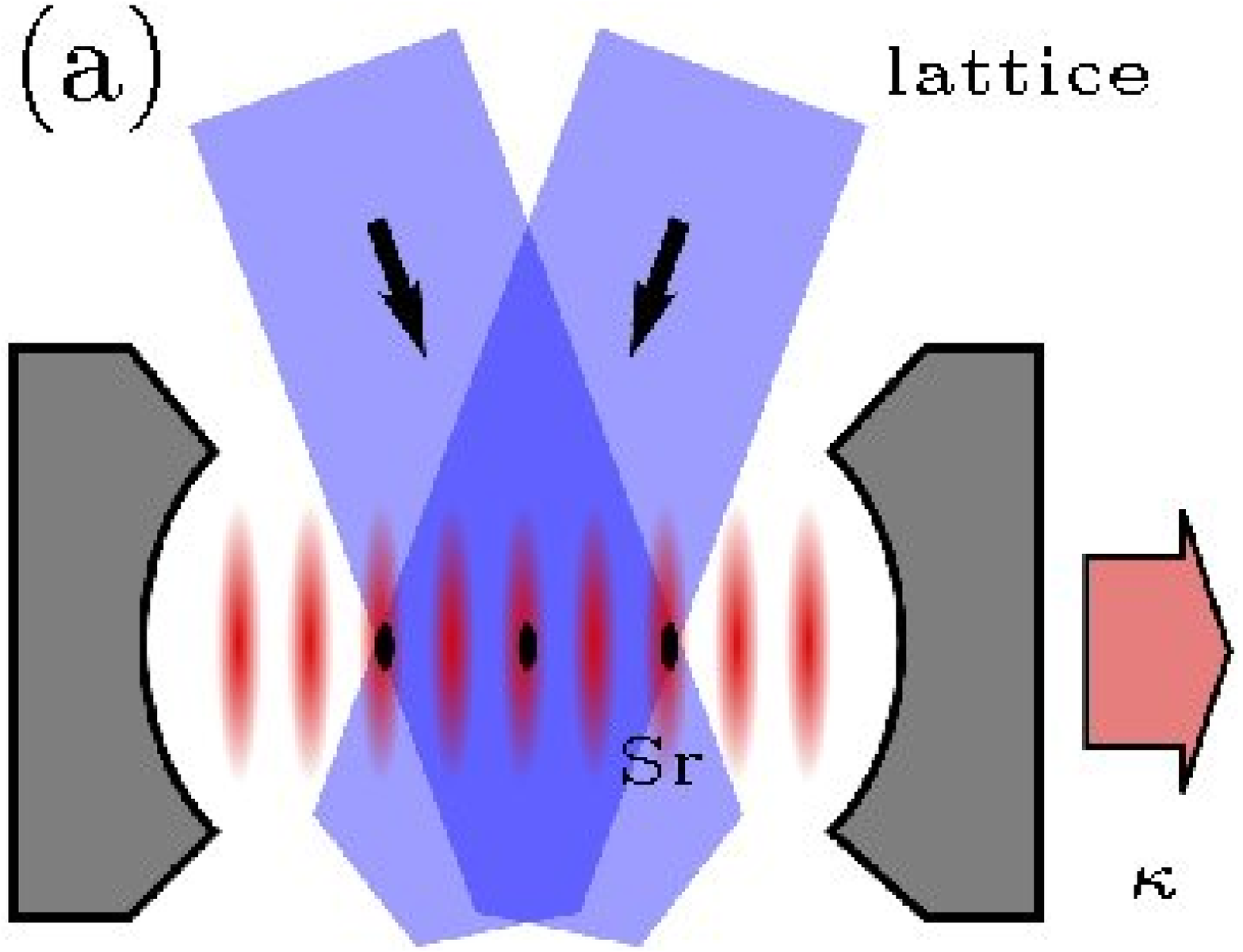}%
  \hspace{\stretch{1}}
  \includegraphics[height=2.9cm]{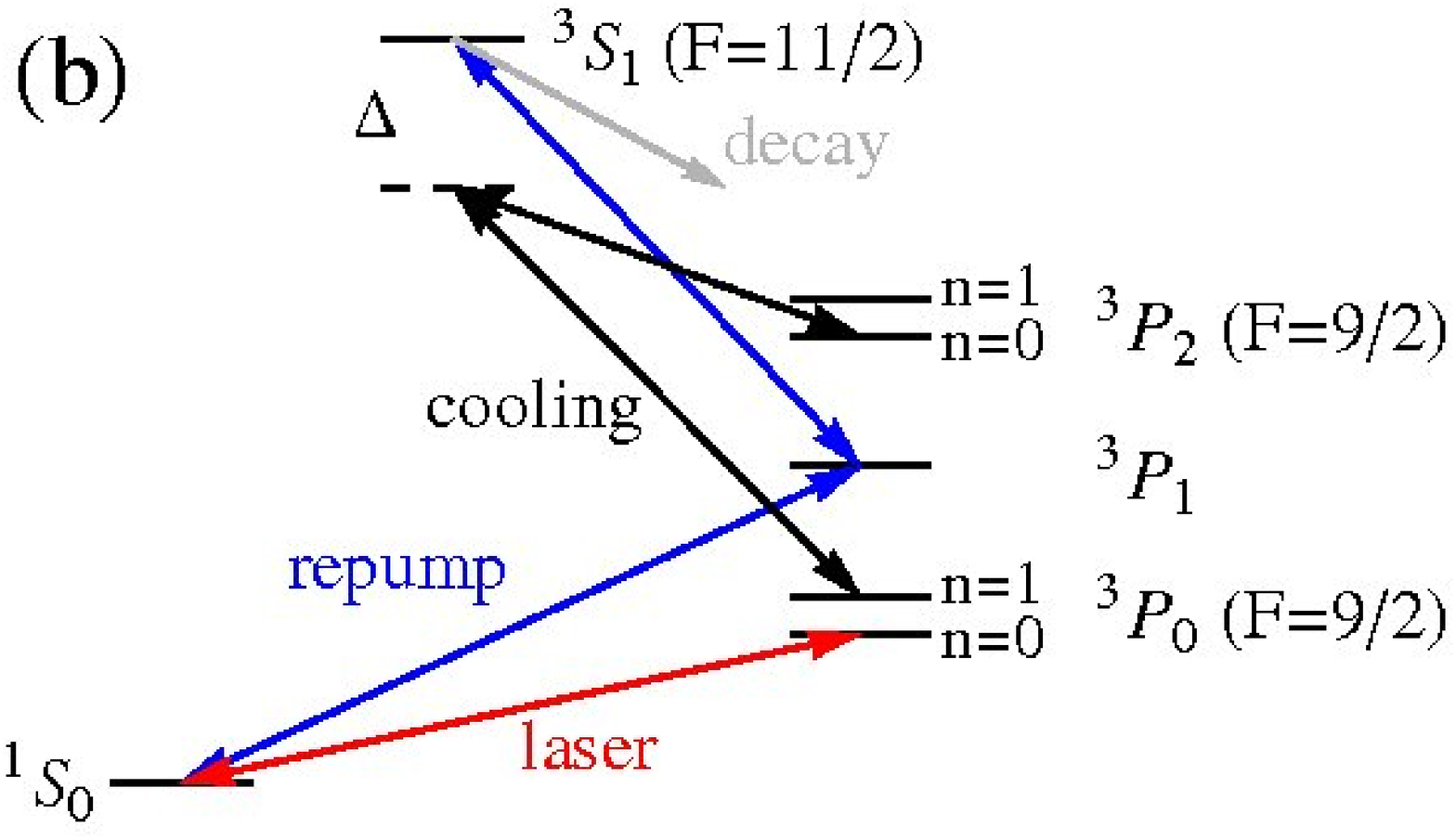}%
  \caption{(Color online) (a) Experimental scheme and (b) the level structure.
  The ultra-narrow ${}^3P_0$--${}^1S_0$ transition indicated by the red arrow
  is the laser transition. The repumping lasers are indicated by the blue
  arrows, spontaneous decay into the ${}^3P$ states by the gray arrow,
  and the Raman side-band cooling lasers by the black arrows. The $n$ quantum
  numbers indicate the vibrational levels of ${}^3P_0$ and ${}^3P_2$.}
  \label{schematic}
\end{figure}

The key observation that motivates this work is that if we could coerce the
ensemble of atoms to emit the energy stored in them collectively rather than
individually, the resulting power of order $10^{-12}$ W would be large enough
to be technologically relevant. We show that such collective emission of
photons can indeed be accomplished if the atoms are located inside a high-Q
cavity. The collective interaction between atoms and cavity fields
\cite{Vuletic:CollectiveCooling,Domokos:CollectiveCooling,Nagorny:Collective2003,Meiser:SpinSqueezing} as well as lasers based on emission from
microscopic atom samples \cite{Meschede:oneatommaser,PhysRevLett.76.1800EtAl,McKeeverEtAl:singleatomlaser,Rice:CavityQEDLaser} are of great theoretical
and experimental interest in quantum optics.

Such a laser would operate in a strikingly different regime from conventional
lasers: the cavity relaxation rate exceeds the atomic relaxation rates by many
orders of magnitude. This system is thus an extreme case of a bad-cavity laser.
Such bad cavity lasers have been studied in the past
\cite{Kuppens:QuantumLimitedBadCavityLaser,KolobovEtAl:PumpingStatisticsFluctuations,Haken:EncyclopediaOfPhysics},
although in those papers the separation of timescales was not nearly as large
as is considered here. In the microwave domain masers are prime examples of a bad cavity laser. Also in the previous systems the cavity field
contains macroscopic numbers of photons while in our system the field
occupation number can be $\ll 1$. The extremity of the
current system makes it necessary that we revisit the foundations of the theory
to obtain reliable predictions. We will focus on the
example of ${}^{87}$Sr confined in a lattice potential to make the presentation
clearer and more concise. Our results however are general and can be easily
translated to other alkaline-earth-like atomic systems with similar structure.

The general setup is shown schematically in Fig.~\ref{schematic}. We consider
$N$ ultra-cold two level atoms with transition frequency $\omega_a$ confined in
an optical lattice. $e$ and $g$ correspond to the two clock
levels, ${}^3P_0$ and ${}^1S_0$ in ${}^{87}$Sr. The atoms have a spontaneous
decay rate $\gamma$ and we model all inhomogeneous processes by an effective
relaxation rate $T_2^{-1}$ for the atomic dipole. The laser transition is
coupled to a single mode of a high $Q$-optical resonator with resonance
frequency $\omega_c$ and linewidth $\kappa$. We assume that the atoms are held
in place by the external optical lattice potential exactly in such a way that
all atoms couple maximally and with the same phase to the specific cavity mode.
Repumping lasers resonantly drive transitions from the ${}^1S_0$ to the
${}^3P_1$ transition and from there to the ${}^3S_1$ state. The repump lasers
optically pump the atoms into the ${}^3P_0$ and ${}^3P_2$ states. A Raman
transition from ${}^3P_0$ to ${}^3P_2$ via ${}^3S_1$ is used to implement
side-band cooling to the vibrational ground state and to optically pump all
atoms to the ${}^3P_0$ state thus providing inversion for the laser transition.
We summarize all these repump steps by an effective repump rate $w$. We have
estimated AC Stark shifts incurred by these repump and cooling lasers and we
find that for repump rates up to around $w\sim 10^3 s^{-1}$ the AC stark shift
induced fluctuations of the atomic transition frequency is negligible at the
mHz level, if the laser intensity is stabilized to 1\%. With the continuous
cooling in place we are justified in assuming that the atoms are in the ground
state of the lattice.

A discussion of the relative and absolute magnitude of the different rate
constants is in order. For ${}^{87}{\rm Sr}$, the linewidth of the doubly
forbidden intercombination line is $\gamma\approx 0.01\; {\rm s}^{-1}$. The
inhomogeneous lifetime in the most recent generation of lattice clock
experiments has been pushed to $T_2\sim 1\; {\rm s}$ \cite{BoydEtAl06Science}.
The effective repump rate $w$ can be widely tuned from $0$ to values of order
$10^4 \;{\rm s}^{-1}$ and beyond. The cavity parameters enter the physics
through the cooperativity parameter $C=\Omega^2/(\kappa\gamma)$ which is
independent of the length of the cavity.  For definiteness we consider a cavity
with an effective mode volume $V_{\rm eff}=(1 {\rm mm})\pi\times(50 \mu {\rm
m})^2$.  Because of the extremely weak dipole matrix element of the inter
combination transition of order $10^{-5}e a_0$, with $e$ the electron charge
and $a_0$ the Bohr radius, this leads to a vacuum Rabi frequency of order
$\Omega\sim 37s^{-1}$. In this setup the cavity decay rate is by far the
largest time scale. For a finesse of $\mathcal{F}=10^6$ and the above cavity
parameters the cavity linewidth is $\kappa=9.4\times 10^5\; {\rm s}^{-1}$, and
thus $C\approx0.15$.

The coupled atom-cavity system can be described by the Hamiltonian
\begin{equation}
  \hat H = \frac{\hbar \omega_a}{2} \sum_{j=1}^{N}\hat
  \sigma_j^z+\hbar\omega_c\hat a^\dagger \hat a+\frac{\hbar \Omega
  }{2}\sum_{j=1}^N\left(\hat a^\dagger  \hat
  \sigma_j^-+{\rm H.c.}
  \right)\;.
\end{equation}
In this formula, $\Omega=\hbar^{-1}\sqrt{\hbar \omega_c/(2 \epsilon_0V_{\rm
eff})}$, $\epsilon_0$ the vacuum permittivity, and $\hat a$ and $\hat
a^\dagger$ are bosonic annihilation and creation operators for photons in the
laser mode. We have introduced Pauli matrices $\hat
\sigma_j^z=\ket{e_j}\bra{e_j}- \ket{g_j}\bra{g_j}$ and $\hat
\sigma_j^-=(\hat\sigma_j^+)^\dagger=\ket{g_j}\bra{e_j}$ for the $j$th atom.

We take the various decay processes into account by means of the usual
Born-Markov master equation for the reduced atom-cavity density matrix $\hat
\rho$,
$d\hat \rho/dt= (i\hbar)^{-1}[\hat H,\hat\rho]+\mathcal{L}[\hat \rho],$
with the Liouvillian $\mathcal{L}[\hat\rho]=\mathcal{L}_{\rm
cavity}[\hat\rho]+\mathcal{L}_{\rm spont.}[\hat\rho]+\mathcal{L}_{\rm
inhom.}[\hat\rho]+\mathcal{L}_{\rm repump}[\hat\rho]$. The Liouvillian for cavity decay
is $\mathcal{L}_{\rm cavity}[\hat \rho]=-\kappa/2(\hat
a^\dagger\hat a\hat \rho+\hat \rho \hat a^\dagger\hat a-2\hat a\hat\rho\hat
a^\dagger)$, the spontaneous decay of the atoms is
described by $\mathcal{L}_{\rm spont}[\hat
\rho]=-\gamma/2\sum_{j=1}^N\left(\hat \sigma_j^+\hat\sigma_j^-\hat \rho+\hat
\rho\hat \sigma_j^+\hat\sigma_j^--2\hat \sigma_j^-\hat
\rho\hat\sigma_j^+\right)$, and the Liouvillian for the inhomogeneous life time
is $\mathcal{L}_{\rm inhom.}[\hat \rho]=1/(2T_2)\sum_{j=1}^N(\hat\sigma_j^z\hat
\rho\hat \sigma_j^z-\hat \rho)$. The Liouvillian for the repumping is identical
to that for the spontaneous decay with the replacements
$\gamma\rightarrow w$, $\hat \sigma_j^-\rightarrow\hat \sigma_j^+$, and $\hat
\sigma_j^+\rightarrow\sigma_j^-$.

An important aspect of this system that is born out by the master equation is
that the coupling of the atoms to the light field is completely collective. The
emission of photons into the cavity acts to correlate the atoms with each other
similar to the case of ideal small sample super-radiance
\cite{Dicke:Superradiance,Gross:Superradiance}, leading to the build-up of a
collective dipole. The locking of the phases of the dipoles of different atoms
gives rise to a macroscopic dipole that radiates more strongly than independent
atoms. The macroscopic dipole is also more robust against noise from decay
processes and repumping, leading to a reduced linewidth.

We have verified that the system settles to steady state much faster than the
anticipated total operation time, provided that the repump rate is not too
close to the laser threshold derived below. For the representative example
parameters used below, the relaxation oscillations decay after a time
$\lesssim1 {\rm s}$ while the total operation time will typically be $>1$
minute. We therefore focus entirely on the steady state behavior in this
Letter. To find the steady state we introduce a cumulant expansion to second
order for the expectation values of system observables \cite{Kubo:GeneralizedCumulantExpansionMethod}. We denote raw
expectation values by $\langle \ldots\rangle$ and cumulant expectation values
by $\langle \ldots\rangle_c$. In our model all expectation values, cumulant or
raw, are completely symmetric with respect to exchange of particles, for
instance $\langle \hat \sigma^z_j\rangle_c =\langle \hat \sigma^z_1\rangle_c$
and $\langle \hat a^\dagger\hat \sigma^-_j\rangle_c=\langle \hat
a^\dagger\hat\sigma^-_1\rangle_c$ for all $j$, $\langle \hat
\sigma^+_i\hat\sigma^-_j\rangle_c=\langle
\hat\sigma^+_1\hat\sigma^-_2\rangle_c$ for all $i\neq j$, etc. In this
formalism where we explicitly keep track of higher order correlations, the
total phase invariance of the system is not broken and we have $\langle \hat
a\rangle=\langle \hat a^\dagger\rangle=\langle \hat \sigma^\pm_1\rangle=0$.
The only non-zero cumulant of the first order is then the inversion $\langle
\hat\sigma^z_1\rangle_c=\langle\hat\sigma^z_1\rangle$. This cumulant couples to
second order cumulants through the atom-field coupling,
\begin{equation}
  \frac{d \langle \hat\sigma^z_1\rangle_c}{d t}=-(w+\gamma)(\langle \hat\sigma^z_1\rangle_c-d_0) + i \Omega\left(\langle \hat a^\dagger\hat\sigma^-_1\rangle_c-\langle \hat a\hat\sigma^+_1\rangle_c\right)\;,
  \label{inversionEquation}
\end{equation}
where $d_0=(w-\gamma)/(w+\gamma)$. The atom-field coherence $\langle \hat
a^\dagger \hat \sigma^-_1\rangle_c$ evolves according to
\begin{eqnarray}
  \frac{d \langle \hat a^\dagger \hat \sigma^-_1\rangle_c}{dt}&=&
  \label{atomFieldCoherenceEquation}
  -\left(\frac{w+\gamma}{2}+\frac{1}{T_2}+\frac{\kappa}{2}-i\delta\right)\langle \hat a^\dagger \hat \sigma^-_1\rangle_c\\
  &&+\frac{i\Omega}{2}\Big[\langle \hat a^\dagger\hat a \hat\sigma^z_1\rangle_c+\langle \hat a^\dagger\hat a\rangle_c\langle \hat\sigma^z_1\rangle_c\nonumber\\
  &&\qquad +\frac{\langle \hat\sigma^z_1\rangle_c+1}{2}+(N-1)\langle \hat \sigma_1^+\hat\sigma_2^-\rangle_c\Big]\nonumber,
\end{eqnarray}
where $\delta=\omega_c-\omega_a$. We have estimated the third order cumulant in
square brackets to be much smaller than the other terms and therefore neglect
it. The second and third terms represent the exchange of energy between cavity
field and a single atom. They are non-collective in nature. The last term
describes the coupling of the atom-field coherence to the collective spin-spin
correlations which locks the relative phase between atoms and field to the
phase of the macroscopic atomic dipole.

The spin-spin correlations evolve according to
\begin{eqnarray}
  \frac{d \langle \hat \sigma_1^+\hat \sigma_2^-\rangle_c}{dt}&=&
  -(w+\gamma+2T_2^{-1})\langle \hat\sigma_1^+\hat\sigma_2^-\rangle_c\\
  &&+\frac{\Omega\langle \hat\sigma^z_1\rangle_c}{2 i}\Big[\langle \hat a^\dagger\hat \sigma^-_1\rangle_c-\langle\hat\sigma^+_1\hat a\rangle_c\Big],\nonumber
\end{eqnarray}
where we have dropped the small third order cumulants of the type $\langle\hat
a^\dagger\hat \sigma^z_1\hat \sigma^-_2\rangle_c$. To close the set of
equations we also need the equation for the mean photon number,
\begin{equation}
  \frac{d\langle \hat a^\dagger\hat a\rangle_c}{dt}=-\kappa \langle \hat a^\dagger\hat a\rangle_c+\frac{N\Omega}{2i}(\langle \hat a^\dagger\hat\sigma^-_1\rangle_c-\langle \hat\sigma^+_1\hat a\rangle_c).
  \label{intensityEquation}
\end{equation}

We consider the steady state of this system for the case $\delta=0$ by setting
the time derivatives in
Eqns.~(\ref{inversionEquation})-(\ref{intensityEquation}) to zero. The
resulting algebraic equations can be solved exactly. Simple approximate results
can be obtained in certain limits on which we will focus in our discussion to
explain the underlying physics. The numerical results reproduced in the figures
are based on the exact solutions and agree well with the approximate treatment.

Our first goal is to understand the role of the collective effects. Neglecting
all decay constants but $\kappa$ in Eq.~(\ref{atomFieldCoherenceEquation}),
keeping only the collective term proportional to $\langle \hat
\sigma_1^+\hat\sigma_2^-\rangle_c$, and approximating $N-1\approx N$ we find
for the atom-field coherence in steady
state $\langle \hat
a^\dagger\hat\sigma^-_1\rangle=iN\Omega\kappa^{-1}\langle\hat
\sigma_1^+\hat\sigma_2^-\rangle_c$. Inserting this result into the steady state
equation for the inversion determines the saturated inversion and plugging that
and $\langle \hat a^\dagger\hat\sigma^-_1\rangle$ into the steady state equation
for the spin-spin correlations yields the central equation
\begin{equation}
  0=\langle \hat \sigma_1^+\hat \sigma_2^-\rangle_c\left(-\Gamma + d_0 N\gamma C-2\frac{N^2\gamma^2C^2}{w+\gamma}\langle \hat \sigma_1^+\hat \sigma_2^-\rangle_c\right)\;,
  \label{symmetryBreakingEquation}
\end{equation}
where $\Gamma=\gamma+w+2/T_2$ is the total relaxation rate of the atomic
dipole. The solution for $\langle \hat \sigma_1^+\hat \sigma_2^-\rangle_c$
corresponding to the term in parenthesis is the physically stable solution. The
laser threshold is the pump rate at which the gain $d_0 N \gamma C$ overcomes
the losses $\Gamma$. In the limit $\Gamma/(\gamma N C)\rightarrow 0$ this
condition turns into $w>\gamma$. At threshold the pump overcomes the atomic
losses which is in contrast to conventional lasers where threshold is obtained
when the pump overcomes the cavity losses.

At threshold the spin-spin correlations change sign signifying the onset of
collective behavior. Interestingly the spin-spin correlations change sign again
at a larger pump rate above which the atoms return to normal non-collective
emission. This upper threshold comes about because $d_0$ eventually saturates
at $1$ while the pump induced noise grows with $w$. Setting $d_0=1$ and
neglecting all atomic noise sources other than $w$ we find for the maximum pump
rate
\begin{equation}
  w_{\rm max}=N C\gamma\;.
  \label{maxPumpRate}
\end{equation}
Above this threshold the pump noise destroys the coherences between different
spins faster than the collective interaction induced by the light field can
establish them.

\begin{figure}
  \includegraphics[width=6cm]{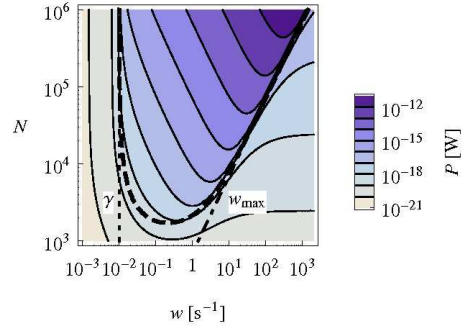}
  \caption{(Color online) Power as a function of pump rate $w$ and atom
  number $N$. The rapid build up of power above threshold $w\sim \gamma$
  can be seen as well as the decrease of emitted power for too strong a
  pump. The dashed line shows the boundary of the region of collective emission
  as determined by the zero of the term in parenthesis in Eq.~(\ref{symmetryBreakingEquation}). Here, $\gamma=0.01 \,{\rm
  s}^{-1}$, $1/T_2=1 \,{\rm s}^{-1}$, $\Omega=37 \,{\rm s}^{-1}$, and $\kappa = 9.4
  \cdot 10^5 \,{\rm s}^{-1}$.}
  \label{intensityPlot}
\end{figure}

A minimum number of particles is necessary to have
collective behavior. Below this critical number, $\langle \hat
\sigma_1^+\hat \sigma_2^-\rangle_c$ is never positive and hence the spin-spin
correlations are never collective. The critical particle number can be estimated
from Eq.~(\ref{symmetryBreakingEquation}). Assuming $T_2\gamma\ll1$, we find
$
  N_{\rm crit} = 2/(C \gamma T_2)
$.
Physically, this equation means that there must be enough particles for the
system to be in the collective strong coupling regime.

The collective versus non-collective behavior is nicely illustrated by the
outcoupled laser power as a function of $w$ and $N$ (Fig.~\ref{intensityPlot}).
Above threshold the outcoupled power rapidly increases until the collective
dipole is destroyed at the second threshold $w_{\rm max}$.

Equation~(\ref{symmetryBreakingEquation}) also allows us to determine the
maximum spin-spin correlation of $\langle \hat \sigma_1^+\hat
\sigma_2^-\rangle_c=1/8$ which is obtained for the pump rate $w_{\rm
opt}= N C\gamma/2$. At this pump rate the laser power reaches its maximum of
\begin{equation}
  P_{\rm max}=\hbar \omega_a N^2 C \gamma/8.
  \label{maxPower}
\end{equation}
The scaling of that power with the square of the number of atoms underlines the
collective nature of the emission. Remarkably, this power is only a factor 2
smaller than the power expected for perfect super-radiant emission from the
maximally collective Dicke state at zero inversion. 
Figure~\ref{intensityPlot} shows that an outcoupled power of order $10^{-12}$~W
is possible with $10^6$ atoms.

From the perspective of potential applications the most striking feature of
this laser is its ultra-narrow linewidth. To find the spectrum we use the
quantum regression theorem to find the equations of motion for the two time
correlation function of the light field $\langle \hat a^\dagger(t)\hat
a(0)\rangle$. This correlation function is coupled to the atom-field
correlation function $\langle \hat \sigma^+(t)\hat a(0)\rangle$. Factorizing
$\langle \hat \sigma^z(t)\hat a^\dagger(t)\hat a(0)\rangle\approx \langle \hat
\sigma^z(t)\rangle\langle\hat a^\dagger(t)\hat a(0)\rangle$ we arrive at the closed set of equations
\begin{equation}
  \frac{d}{dt}
  \left[
  \begin{array}{c}
    \langle \hat a^\dagger(t)\hat a(0)\rangle\\
    \langle \hat \sigma^+(t)\hat a(0)\rangle
  \end{array}
  \right]=
  \left[
  \begin{array}{cc}
    -\kappa/2&\frac{iN\Omega}{2}\\
    -\frac{i\Omega\langle \hat \sigma^z\rangle_c}{2}&-\Gamma/2
  \end{array}
  \right]
  \left[
  \begin{array}{c}
    \langle \hat a^\dagger(t)\hat a(0)\rangle\\
    \langle \hat \sigma^+(t)\hat a(0)\rangle
  \end{array}
  \right]\;.
  \label{qrtEquations}
\end{equation}
The initial conditions are the steady state solutions discussed above. Laplace
transforming the solutions yields the spectrum which is Lorentzian with
linewidth $\Delta \nu$ for $\omega_a=\omega_c$ .

\begin{figure}
  \includegraphics[width=8cm]{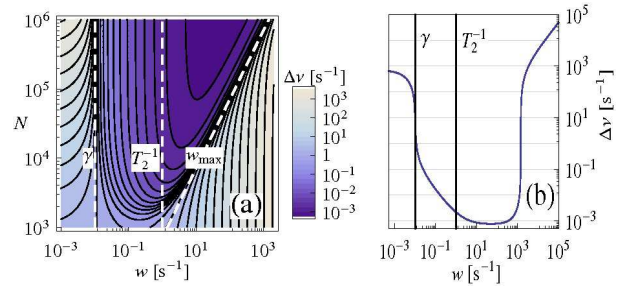}
  \caption{(Color online) (a) Linewidth vs. $w$ and
  $N$. The white dashed lines indicate (from left to right) the
  spontaneous decay rate $\gamma$, the inhomogeneous relaxation rate $1/T_2$,
  and the maximum pump rate $w_{\rm max}$ (Eq.~(\ref{maxPumpRate})).
  Parameters as in Fig.~\ref{intensityPlot}. (b) is a cut through figure (a)
  for $N=10^6$ atoms.}
  \label{lineWidthPlot}
\end{figure}

For our example parameters $\Delta \nu$ is shown in Fig.~\ref{lineWidthPlot}.
The leftmost dashed line in that figure is $\gamma$ corresponding
to the threshold for collective behavior. When the pump strength $w$ passes
through that threshold the linewidth gets rapidly smaller with increasing $w$.
When $w$ reaches $1/T_2$, indicated by the second dashed line, essentially all
atoms are phase locked together. From that point on the pump noise due to $w$
grows in proportion to the size of the collective spin vector. Therefore the
linewidth is approximately constant. Making similar approximations as in the
steady state calculations we obtain the estimate for the minimum laser
linewidth,
$
  \Delta \nu=C \gamma
$.
That estimate agrees well with our numerical results. It is important to note
that the parameters for achieving the maximum outcoupled power
Eq.~(\ref{maxPower}) and the minimum linewidth are compatible with each other.
For our example parameters the estimate yields a linewidth smaller than the
homogeneous linewidth of the atomic clock transition. We have studied the
dependence of the laser frequency on the atom-cavity detuning and we find that
this narrow linewidth can be observed if the cavity resonance frequency is
stable at the $1$ kHz level which is relatively easy to achieve experimentally.
When $w$ increases beyond $w_{\rm max}$, indicated by the third dashed
line, the collective dipole is destroyed and the linewidth increases rapidly
until it is eventually given by $w$. 

Future research is targeted at fully understanding the recoil effects and  
the detailed nature of the joint atomic and field state.

We gratefully acknowledge stimulating discussions with D. Kleppner, J. K. Thompson, and J. Cooper. This work was supported by DFG, DARPA, NIST, DOE, and NSF.

\bibliographystyle{apsrev}
\bibliography{mybibliographyEtAl}

\end{document}